\documentclass[]{raa}            % referee version: for submission
\usepackage{graphicx,times}
\usepackage{natbib}
\usepackage{amssymb}
\usepackage{ulem}
\usepackage{color}

\providecommand{\bm}[1]{\mbox{\boldmath$#1$\unboldmath}}

\begin{document}

   \title{Two particular EA-type binaries in the globular cluster $\omega$ Centauri}

 \volnopage{ {\bf 2013} Vol.\ {\bf X} No. {\bf XX}, 000--000}
   \setcounter{page}{1}

   \author{K. Li
      \inst{1,2,3}
   \and S.-B. Qian
      \inst{1,2,3}
   }

   \institute{National Astronomical Observatories/Yunnan Observatory,
Chinese Academy of Sciences, P. O. Box 110, 650011 Kunming, China; {\it e-mail: likai@ynao.ac.cn} \\
        \and
             Key Laboratory for the Structure and Evolution of Celestial
Objects, Chinese Academy of Sciences
        \and
             University of the Chinese Academy of Sciences, Beijing 100049, China
% \email{likai@ynao.ac.cn}
\vs \no
%  {\small Received [year] [month] [day]; accepted [year] [month] [day] }
}

\abstract{We analyzed the $B$ and $V$ light curves of two EA-type binaries V211 and NV358 using the WD code for the first time. Our analysis shows that V211 is a typical Algol-type binary and NV358 is a well detached binary system. As the two binaries are definite proper motion members of $\omega$ Centauri, we estimated their physical parameters, obtaining $M_1=1.13\pm0.03M_\odot,$ $R_1=0.98\pm0.01R_\odot$ and $M_2=0.33\pm0.01M_\odot,$ $R_2=0.92\pm0.01R_\odot$ for V211; $M_1=1.30\pm0.05M_\odot,$ $ R_1=1.03\pm0.01R_\odot$ and $M_2=0.58\pm0.02M_\odot,$ $R_2=0.78\pm0.01R_\odot$ for NV358. On the color-magnitude diagram of $\omega$ Centauri, V211 is located in the faint blue straggler region and its primary component is more massive than a star at the main-sequence turnoff. Therefore, V211 is a blue straggler and should be formed by mass transfer from the secondary component to the primary. The age of NV358 is less than 1.93 Gyr, indicating that it is much younger than first-generation stars in $\omega$ Centauri. As NV364 in $\omega$ Centauri, NV358 should be a second-generation binary.
 }

   \authorrunning{Li and Qian }    %author_head in even pages
   \titlerunning{ Two particular EA-type binaries in the globular cluster $\omega$ Centauri }  % title_head in odd pages
   \maketitle
\section{Introduction}           %% first-level sections will be auto-capitalized
\label{sect:intro}
Eclipsing binaries are very rare in globular clusters, but they play a significant role in the dynamical evolution of globular clusters and stellar populations studies. They provide the source of energy that can against and avoid core collapse in globular clusters (\citealt{goo89}). They are also particularly interesting. Some Algol type eclipsing binaries such as NJL5, V239 in $\omega$ Centauri and V228 in 47 Tuc (\citealt{hel93,li12a,kal07}) have been identified to be blue stragglers (BSs). They are good tools to comprehend the BS formation theories. Detached double line spectroscopic eclipsing binaries in globular clusters provide the opportunity to calculate the age of the cluster based on their well determined physical properties of mass and radius (see \citealt{tho01,tho10}).

BS stars appear to be anomalously younger than other stars of their population. In the color-magnitude diagrams (CMDs) of clusters, these stars are bluer and brighter than the main-sequence turnoff. They were first noticed in the globular cluster M3 by \cite{san53}. Now, not only in globular and open clusters, dwarf galaxies (\citealt{pio04,de06,mom07}), but also in the field (\citealt{car05}), investigators have discovered BSs. The formation of BSs is still controversial. Several mechanisms have been proposed to explain BS formation. The two most popular BS formation mechanisms are mass transfer in close binary systems (\citealt{mcc64,car01}) and direct collision between two stars (\citealt{hil76}). \cite{per09} discussed another possibility of BS formation in primordial (or dynamically) hierarchical triple star systems. By studying this type of objects, one can reveal the dynamical history of a cluster and the role of dynamics on the stellar evolution. BSs statistics can also provide some constrains for initial binary properties. The bimodal BS radial distribution in many globular clusters (e.g. \citealt{fer04,map06,lan07,bec11}) has been observed.
A scenario that BSs in the dense core were formed in collisions, whereas BSs in the low density cluster outskirts were formed by mass transfer in close binaries has been suggested to explain this.

The traditional opinion about globular clusters is that all stars within a globular cluster are thought to share the same age and initially homogeneous chemical composition and they are simple stellar populations. This has one noticeable exception: $\omega$ Cen, the most massive cluster in the Milky Way, whose stars show a large spread in metallicity. At present, the situation has been much more complex, and it is now identifiable that almost all the globular clusters so far examined in detail have at least two stellar generations. Clear evidence for multiple main sequences (\citealt{bed04}) and giant branches (\citealt{nat11}), and unusual horizontal branch (\citealt{dan05}) and subgiant branch (\citealt{mil08,mor09}) morphologies all can be explained in straightforward ways by the presence of multiple generations of stars.
The first second-generation binary named NV364 in $\omega$ Centauri has been identified by \cite{li12}.

$\omega$ Centauri is one of the most metal poor globular clusters in the Milky Way, with [Fe/H]=-1.53, and has an interstellar reddening $E(B-V)=0.12$ and a distance modulus $(m-M)_v=13.94$ (\citealt{har96})). V211, with a period of 0.576235 $d$, was first identified in the outskirts of the cluster by \cite{kal96} during a search for variable stars in the central part of the globular cluster $\omega$ Centauri.
NV358, with a period of 0.59964 $d$, was first discovered in the outer region of the cluster by \cite{kal04} during a photometric survey for variable stars in the field of this cluster. On the CDM of the cluster, V211 is located in the faint BS region, while NV358 occupies a position of the bright BS domain. Light curves of several eclipsing binaries in $\omega$ Centauri have been analyzed, but except for the two. In this paper, we show the investigation of $B$ and $V$ light curves of the two binaries taken from \cite{kal04} using the Wilson-Devinney code.

\section{Light curve analysis of the two binaries}
\label{sect:Lig}

$\omega$ Centauri was observed using the 1.0-m Swope telescope at Las Campanas Observatory by \cite{kal04} under The Cluster AgeS Experiment (CASE) project during the interval from 1999 February 6/7 to 2000 August 9/10. $B$ and $V$ light curves of 301 variables were obtained and the photometric data are available on the VizieR Result page. As no photometric analysis has been carried out for the two binaries, V211 and NV358, which have good enough photometric data and are located in the BS region on the CMD of $\omega$ Centauri, we chose them to do further analysis. Using the fourth version of the W-D program (\citealt{wil71,wil90,wil94,wil03}) which is a good tool for the modeling of eclipsing binaries based on real photometric and spectroscopic (radial velocity) data, we analyzed the $B$ and $V$ light curves of the two binaries taken from \cite{kal04} for the first time. Some of the photometric data that have serious derivation from the phased light curves (most in $B$ band) were deleted. The ephemerides used to calculate the phases of the two binaries were
\begin{equation}
Min.I=2451283.7791+0.576235E,
\end{equation}
\begin{equation}
Min.I=2451284.1098+0.599640E.
\end{equation}

During our solutions, the effective temperature of the primary component, $T_1$, was determined based on the values of dereddened $B-V$ at the secondary minimum of the two binaries described below. The gravity-darkening coefficients of the components were taken to be 1.0 for radiative atmosphere ($T\geq7200$) from \cite{von24} and 0.32 for convective atmosphere ($T<7200$) from \cite{luc67}. The bolometric albedo coefficients of the components were fixed at 1.0 and 0.5 for radiative and convective atmospheres following \cite{ruc69}. The bolometric and bandpass limb-darkening coefficients of the components were also fixed (\citealt{van93}). Starting with the solutions by mode 2, we found that the solutions of V211 usually converged when the secondary component fills its Roche lobe and that of NV358 quickly converged. So, the final iterations of V211 were made in mode 5, which corresponds to semi-detached configuration, while that of NV358 was made in mode 2. The quantities varied in the solutions of the two stars were the mass ratio $q$, the effective temperature of the secondary component $T_2$, the monochromatic luminosity of primary component in $B$ and $V$ bands $L_1$, the orbital inclination $i$ and the dimensionless potential of the primary component $\Omega_1$. The dimensionless potential of the secondary component $\Omega_2$ is also a variable for NV358.

\subsection{V211}
The temperature of the primary component, $T_1$, was determined from the dereddened color index $(B-V)_1$ using the program provided by \cite{wor11}. \cite{bel09} determined the membership probability of V211 as 90\%.
We adopted an interstellar reddening of $E(B-V)=0.12$ and a metallicity value of [Fe/H] = -1.53 (\citealt{har96})) for $\omega$ Centauri. The color index at the secondary minimum is a good approximation to the color index of the primary star. Based on the phased data, the color index at the secondary minimum is measured to be $(B-V)_1=0.393$, leading to $T_1=7035K$. The bolometric albedo and the gravity-darkening coefficients of the components were set as $A_1=A_2=0.5$ and $g_1=g_2=0.32$ for convective atmosphere. Bolometric and bandpass square-root limb-darkening parameters of the components taken from \cite{van93} were listed in Table 1. A $q$-search method was used to determine the mass ratio of V211. Solutions were carried out for a series of values of the mass ratio (0.2, 0.25, 0.3, 0.4, 0.5, 0.6, 0.7). The relation between the resulting sum $\Sigma$ of weighted square deviations and $q$ is plotted in Fig. 1. The minimum value was obtained at $q=0.30$. Therefore, we fixed the initial value of mass ratio $q$ at 0.30 and made it an adjustable parameter. Then, we executed a differential correction until it converged and final solutions were derived. The final photometric solutions are listed in Table 1. The comparison between observed and the theoretical light curves is shown in Fig. 1.

\subsection{NV358}
The color index of NV358 at the secondary minimum is 0.142. \cite{bel09} determined the membership probability of NV358 is 99\%. The $(B-V)_{1,0}$  of the primary was fixed at 0.022. Using the same method of V211, we fixed the effective temperature of the primary component of NV358 at $T_1=8918K$. The bolometric albedo and the gravity-darkening coefficients of the components were set to $A_1=A_2=1.0$ and $g_1=g_2=1.0$ for radiative atmosphere. Bolometric and bandpass square-root limb-darkening parameters of the components taken from \cite{van93} were listed in Table 2. A $q$-search method was also used to determine the mass ratio of NV358. The final photometric solutions are listed in Table 2. Fig. 2 shows the comparison between observed and the theoretical light curves.

Because \cite{joh09} found large metallicity spread for stars in the cluster $\omega$ Centauri, a metallicity value of [Fe/H] = -1.0 was also used to determine the effective temperature of the primary component, and $T_1=7104K$ for V211 and $T_1=8923K$ for NV358 were obtained. The solutions based on the metallicity of [Fe/H] = -1.0 are listed in Tables 1 and 2. We find that the additional solution results of this metallicity value are in accordance to previous solutions. Therefore, the results using [Fe/H] = -1.53 are adopted to be the final solution.

\begin{figure*}
\begin{center}
\includegraphics[width=0.5\textwidth]{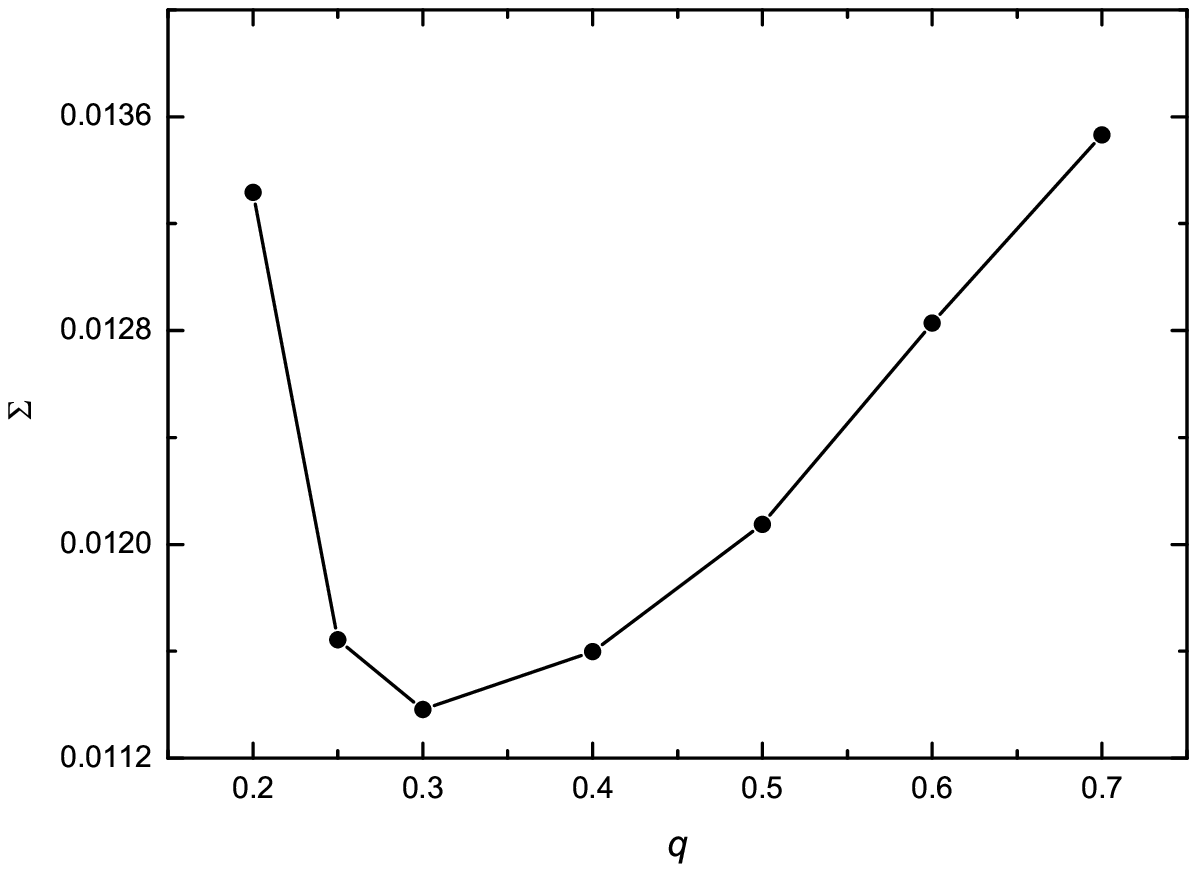}%
\includegraphics[width=0.5\textwidth]{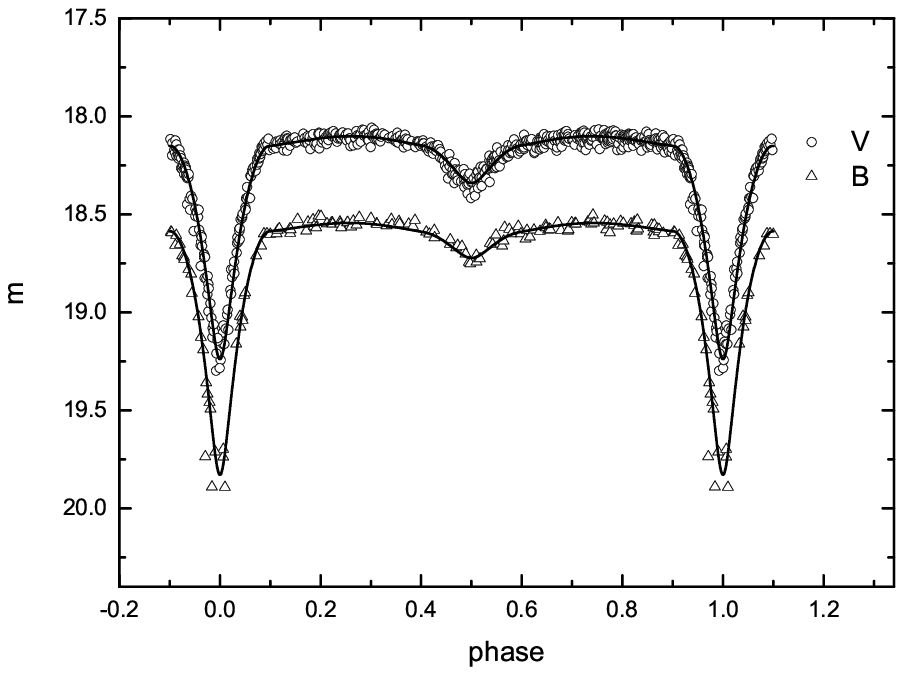}

\caption{Left panel shows $q$-search for V211. Right panel displays observed (open symbols) and theoretical (solid lines) light curves of V211 in $BV$ passbands.}
\end{center}
\end{figure*}
\begin{figure*}
\begin{center}
\includegraphics[width=0.5\textwidth]{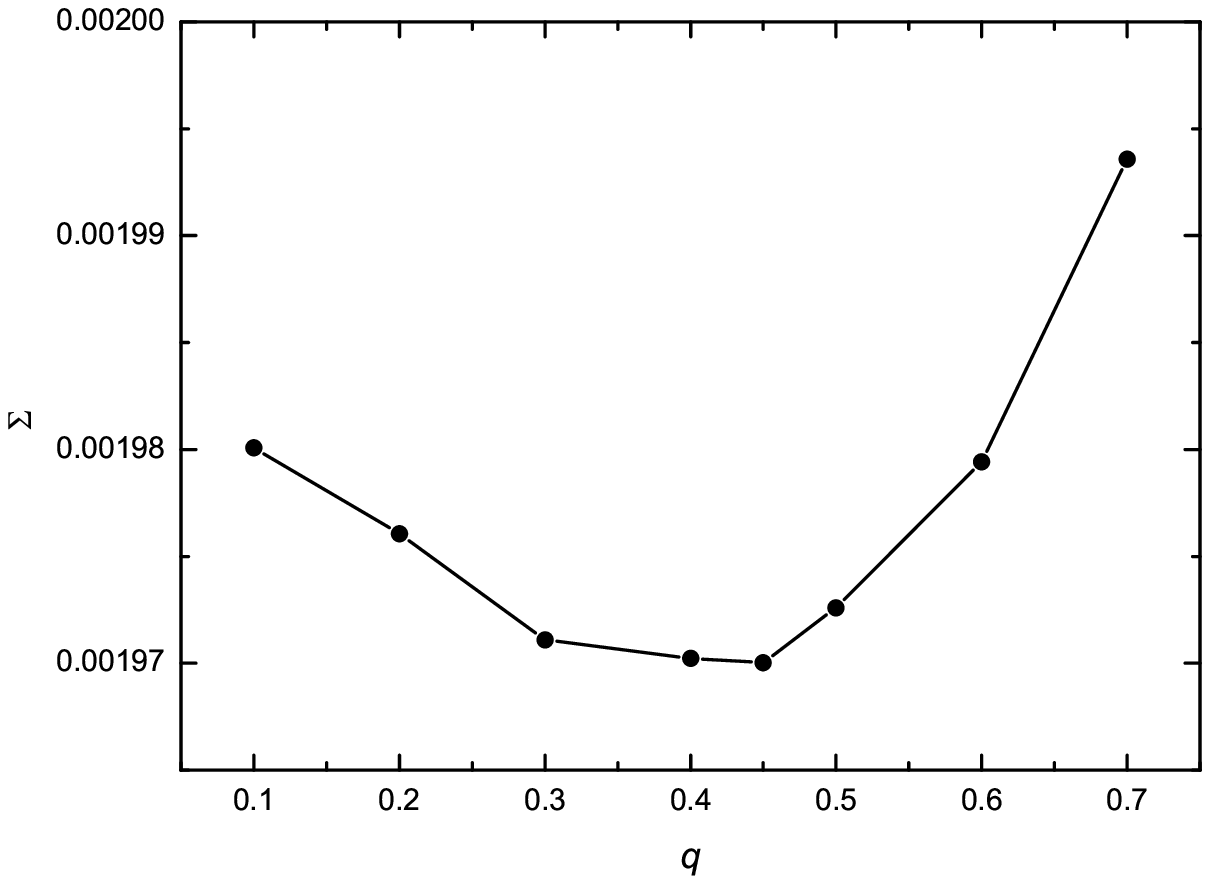}%
\includegraphics[width=0.5\textwidth]{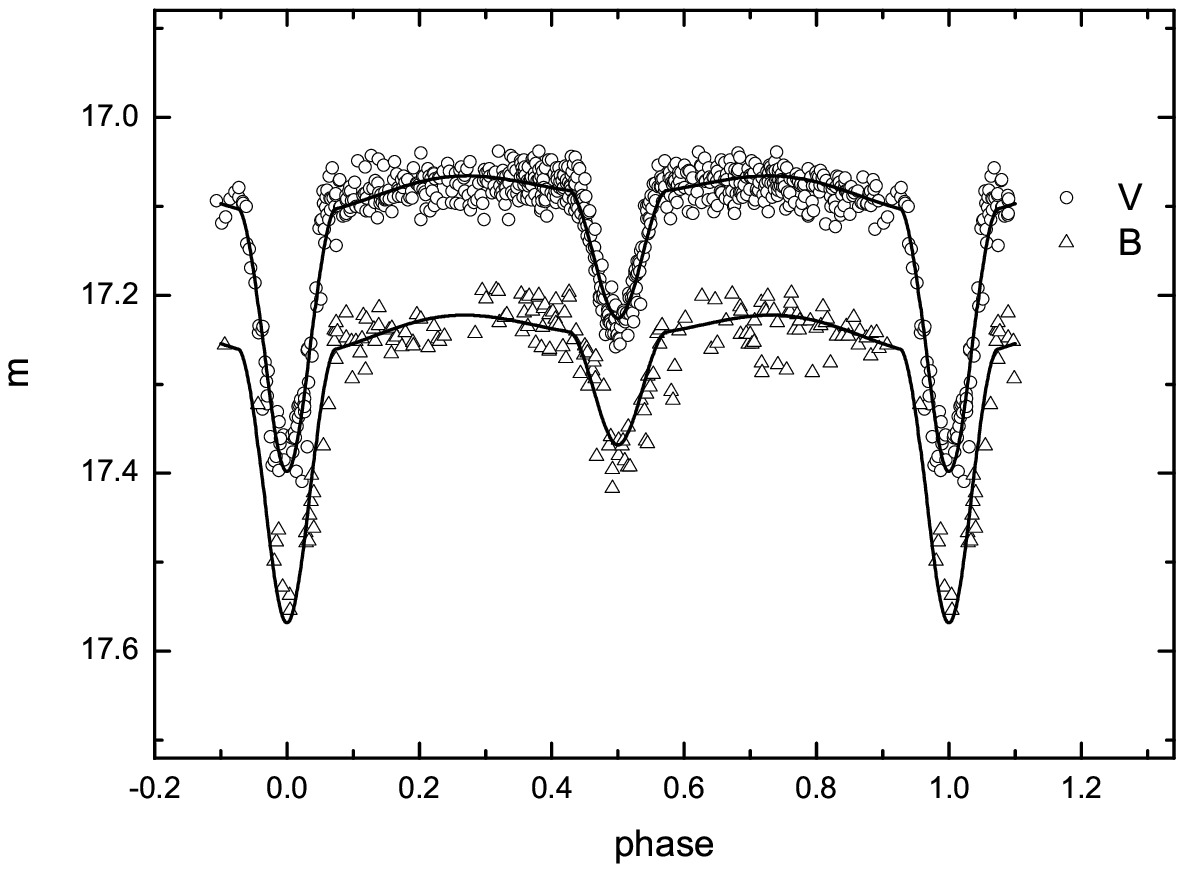}

\caption{Left panel shows $q$-search for NV358. Right panel displays observed (open symbols) and theoretical (solid lines) light curves of NV358 in $BV$ passbands.}
\end{center}
\end{figure*}

\begin{table*}
\begin{center}
\caption{Photometric solutions for V211 in the globular cluster $\omega$ Centauri}
\begin{tabular}{lclcl}
\hline
Parameters           & [Fe/H] = -1.53 & Errors& [Fe/H] =  -1.0& Errors \\
\hline

    $ g_1=g_2$       &       0.32 & Assumed          &      0.32 & Assumed          \\
     $A_1=A_2$       &       0.5 & Assumed          &       0.5 & Assumed          \\
     $x_{1bol}$      & 0.093& Assumed               & 0.093& Assumed               \\
     $x_{2bol}$      & 0.265& Assumed               & 0.265& Assumed               \\
     $y_{1bol}$      & 0.632& Assumed               & 0.632& Assumed               \\
     $y_{2bol}$      & 0.441& Assumed               & 0.441& Assumed               \\
      $x_{1B}$       & 0.173& Assumed               & 0.173& Assumed               \\
     $x_{2B}$        & 0.757& Assumed               & 0.757& Assumed               \\
     $y_{1B}$        & 0.706& Assumed               & 0.706& Assumed               \\
     $y_{2B}$        & 0.109& Assumed               & 0.109& Assumed               \\
     $x_{1V}$        & 0.059& Assumed               & 0.059& Assumed               \\
     $x_{2V}$        & 0.445& Assumed               & 0.445& Assumed               \\
     $y_{1V}$        & 0.726& Assumed               & 0.726& Assumed               \\
     $y_{2V}$        & 0.398& Assumed               & 0.398& Assumed               \\
    $ T_1(K) $       &       7035 & Assumed         &       7104 & Assumed         \\

  $q(M_2/M_1) $      &     0.2941&  $ \pm0.0094$      &     0.2942&  $ \pm0.0093$      \\

        $T_2(K) $    &       5219 & $ \pm24$        &       5257 & $ \pm24$        \\

         $i$         &     85.166 & $ \pm0.416$     &     85.172 & $ \pm0.415$     \\

$L_1/(L_1+L_2)(B) $  &      0.8739 & $ \pm0.0008$   &      0.8737 & $ \pm0.0008$   \\
$L_1/(L_1+L_2)(V) $  &      0.8160 & $ \pm0.0013$   &      0.8160 & $ \pm0.0013$   \\

        $\Omega_1$   &    3.7041 & $ \pm0.0562$     &    3.7055 & $ \pm0.0566$     \\

        $\Omega_2$   &    2.4532 & Assumed     &    2.4534 &  Assumed     \\

  $r_1(pole)$        &   0.2922 &$ \pm 0.0048$      &   0.2921 &$ \pm 0.0049$      \\

  $r_1(side)$        &   0.2972 & $ \pm0.0052$      &   0.2971 & $ \pm0.0052$      \\

  $r_1(back)$        &     0.3002 & $ \pm0.0054$    &     0.3001 & $ \pm0.0054$    \\

  $r_2(pole)$        &     0.2597 & $ \pm0.0023$    &     0.2596 & $ \pm0.0023$    \\

  $r_2(side)$        &   0.2704 & $ \pm0.0024$      &   0.2704 & $ \pm0.0024$      \\

  $r_2(back)$        &   0.3031 & $ \pm0.0024$      &   0.3031 & $ \pm0.0024$      \\
\hline
\end{tabular}
\end{center}
\end{table*}

\begin{table*}
\begin{center}
\caption{Photometric solutions for NV358 in the globular cluster $\omega$ Centauri}
\begin{tabular}{lclcl}
\hline
Parameters           & [Fe/H] = -1.53 & Errors& [Fe/H] =  -1.0& Errors \\
\hline

    $ g_1=g_2$       &       1.0 & Assumed          &       1.0 & Assumed          \\
     $A_1=A_2$       &       1.0 & Assumed          &       1.0 & Assumed          \\
     $x_{1bol}$      & 0.435& Assumed               & 0.435& Assumed               \\
     $x_{2bol}$      & 0.082& Assumed               & 0.082& Assumed               \\
     $y_{1bol}$      & 0.254& Assumed               & 0.254& Assumed               \\
     $y_{2bol}$      & 0.646& Assumed               & 0.646& Assumed               \\
      $x_{1B}$       & 0.061& Assumed               & 0.061& Assumed               \\
     $x_{2B}$        & 0.125& Assumed               & 0.125& Assumed               \\
     $y_{1B}$        & 0.792& Assumed               & 0.792& Assumed               \\
     $y_{2B}$        & 0.753& Assumed               & 0.753& Assumed               \\
     $x_{1V}$        & 0.039& Assumed               & 0.039& Assumed               \\
     $x_{2V}$        & 0.036& Assumed               & 0.036& Assumed               \\
     $y_{1V}$        & 0.701& Assumed               & 0.701& Assumed               \\
     $y_{2V}$        & 0.743& Assumed               & 0.743& Assumed               \\
    $ T_1(K) $       &       8918 & Assumed         &       8923 & Assumed         \\

  $q(M_2/M_1) $      &     0.4442&  $ \pm0.0125$      &     0.4443&  $ \pm0.0125$      \\

        $T_2(K) $    &       7249 & $ \pm47$        &       7255 & $ \pm48$        \\

         $i$         &     76.033 & $ \pm0.272$     &     76.037 & $ \pm0.272$     \\

$L_1/(L_1+L_2)(B) $  &      0.8247 & $ \pm0.0016$   &      0.8248 & $ \pm0.0016$   \\
$L_1/(L_1+L_2)(V) $  &      0.7873 & $ \pm0.0019$   &      0.7875 & $ \pm0.0020$   \\
      $\Omega_{crit}$&    2.7669 & Assumed        &    2.7672 & Assumed\\
        $\Omega_1$   &    4.0500 & $ \pm0.0447$     &    4.0463 & $ \pm0.0448$     \\

        $\Omega_2$   &    3.4006 & $ \pm0.0607$     &    3.4027 & $ \pm0.0607$     \\

  $r_1(pole)$        &   0.2761 &$ \pm 0.0035$      &   0.2764 &$ \pm 0.0035$      \\

  $r_1(point)$        &   0.2861 &$ \pm 0.0041$      &   0.2864 &$ \pm 0.0041$      \\

  $r_1(side)$        &   0.2805 & $ \pm0.0037$      &   0.2808 & $ \pm0.0038$      \\

  $r_1(back)$        &     0.2841 & $ \pm0.0039$    &     0.2845 & $ \pm0.0040$    \\

  $r_2(pole)$        &     0.2072 & $ \pm0.0063$    &     0.2071 & $ \pm0.0063$    \\

  $r_2(point)$        &   0.2192 &$ \pm 0.0082$      &   0.2190 &$ \pm 0.0082$      \\

  $r_2(side)$        &   0.2103 & $ \pm0.0067$      &   0.2101 & $ \pm0.0067$      \\

  $r_2(back)$        &   0.2166 & $ \pm0.0077$      &   0.2165 & $ \pm0.0077$      \\
\hline
\end{tabular}
\end{center}
\end{table*}

\section{Results and Discussions }
Based on the $B$ and $V$ light curves, photometric solutions for the two EA-type binaries, V211 and NV358, have been derived. It is shown that V211 is a typical Algol-type binary and NV358 is a well detached binary system, the primary and secondary components of NV358 fill 68.3\% and 81.4\% of their critical Roche lobes, respectively. Using the fractional luminosities of the components from the $B$ and $V$ light-curve solutions, quoted above, and adopting
the observed $B$ and $V$ magnitudes of the two binaries at maximum light, we found the following visual magnitudes of the components of the two binaries: $V_1=18.331\pm0.002$, $B_1=18.686\pm0.001$, $V_2=19.948\pm0.008$ and $B_2=20.788\pm0.007$ for V211, $V_1=17.330\pm0.003$, $B_1=17.449\pm0.002$, $V_2=18.751\pm0.010$ and $B_2=19.131\pm0.010$ for NV358. The mean source of the errors is the respective uncertainties in the solutions. Fig. 3 shows the positions of the individual eclipsing components on the CMD (\citealt{nob91}) of $\omega$ Centauri. V211 is located in the faint BS region, while NV358 occupies a position in the bright BS domain.

\subsection{Physical parameters of the two binaries}
According to WFI@2.2m proper-motion catalog of the globular cluster $\omega$ Centauri conducted by \cite{bel09}, V211 (designation 298164) and  NV358 (designation 129801) are definite proper-motion members of the cluster with respective probabilities of 90\% and 99\%. Using the same method as \cite{liu11}, we estimated the physical parameters of the two binaries by the light-curve program of the WD code. First, we calculated the absolute bolometric magnitudes of the two binaries based on the $V$-band absolute magnitudes of them. Second, using the light-curve program of the W-D code, we could also obtain the absolute bolometric magnitudes of the two binaries, which can be compared to the results in the previous step. When the results of the two steps are consistent with each other, the physical parameters of the two binaries are obtained. The physical parameters of the two binaries are listed in Table 3, where the errors represent the uncertainty of $q$.
The mean physical parameters of the two binaries are as follows: for V211, $M_1=1.13\pm0.03M_\odot,$ $R_1=0.98\pm0.01R_\odot$ and $M_2=0.33\pm0.01M_\odot,$ $R_2=0.92\pm0.01R_\odot$; for NV358, $M_1=1.30\pm0.05M_\odot,$ $R_1=1.03\pm0.01R_\odot$ and $M_2=0.58\pm0.02M_\odot,$ $R_2=0.78\pm0.01R_\odot$.
\begin{table}
\begin{center}
\caption{Physical parameters of V211 and NV358}
\begin{tabular}{lclcl}
\hline
 Parameters &      Values & Errors &      Values & Errors\\
 &V211&&NV358&\\
 \hline

    $M_1$ ($M_\odot$)&       1.13 & $\pm0.03$ &       1.30 & $\pm0.05$\\

    $M_2$ ($M_\odot$)&       0.33 & $\pm0.01$&       0.58 & $\pm0.02$ \\

    $R_1$ ($R_\odot$)&        0.98 & $\pm0.01$ &        1.03 & $\pm0.01$\\

    $R_2$ ($R_\odot$) &       0.92 &$\pm0.01$&       0.78 &$\pm0.01$ \\

      $A$ ($R_\odot$) &       3.30 & $\pm0.02$&       3.68 & $\pm0.04$\\

    $L_{bol1}$ ($L_\odot$) &       2.03 & $\pm0.05$&       5.86 & $\pm0.13$\\

    $L_{bol2}$ ($L_\odot$) &       0.54 & $\pm0.01$&       1.46 & $\pm0.13$\\

$\log{g_1}$ ($cgs$)&       4.51 &  $\pm0.01$&       4.52 &  $\pm0.01$    \\

$\log{g_2}$ ($cgs$)&       4.03 & $\pm0.01$&       4.42 & $\pm0.02$     \\

$M_{bol1}$  ($mag$)&       3.98 &  $\pm0.01$&       2.83 &  $\pm0.01$    \\

 $M_{bol2}$ ($mag$)&       5.41 &  $\pm0.01$&       4.34 &  $\pm0.04$    \\

 $M_{bol}$ ($mag$)&     3.722 &   $\pm0.010$&     2.589 &   $\pm0.016$   \\

$m_{v_{max}}$  ($mag$)&      18.11 & Assumed&      17.07 & Assumed     \\
\hline
\end{tabular}
\end{center}
\end{table}

\subsection{The particularity of the two binaries}
Both of the two binaries are very interesting targets. V211 is located in the faint BS region on the CMD of $\omega$ Centauri and is a definite member star of this cluster. Therefore, V211 is an eclipsing BS. The mass of the primary component of V211 is $1.13M_\odot$, which is larger than that of a star at the main sequence turnoff ($M_{TO}=0.8M_\odot$). Liking other eclipsing Algol BS such as NJL5, V239 in $\omega$ Centauri and V228 in 47 Tuc (\citealt{hel93,li12a,kal07}), V211 should be formed by mass transfer from the secondary component to the primary inducing a reversal of the original mass ratio so that the current primary was originally the less massive component. NV358 occupies a position in the bright BS domain of the CMD of $\omega$ Centauri, and the two components of NV358 also stay in the BS region. Both are bluer than the main-sequence stars in $\omega$ Centauri, indicating that they are younger and metal richer. We have derived that the surface gravity of the primary component is $\log{g}=4.52$ $cgs$, suggesting that it is a main-sequence star. Then, we can estimate its age using the equation
\begin{equation}
t_{MS}=\frac{3.37\times10^9}{(M/M_\odot)^{2.122}},
\end{equation}
derived from \cite{yil11}. An age of $T_a\leq1.93$ Gyr is obtained for the primary component. It is believed that short period binary systems are formed by a fragmentation process and are unlikely formed from a capture process. Therefore, the components of NV358 should be of the same age. We adopt a value of 1.93 Gyr for the age of NV358. The age of $\omega$ Centauri is $16\pm3$ Gyr (\citealt{nob91}), and almost at the same time the first-generation stars in $\omega$ Centauri formed. Therefore, NV358 is much younger than the first-generation stars in $\omega$ Centauri. As NV364 in $\omega$ Centauri (\citealt{li12}), we deduce that NV358 is also a second-generation binary.

In summary, V211 is an eclipsing BS, making it an important object to comprehend the hypothesis that BSs are formed by mass transfer in close binaries. At the same time, V211 was discovered in the out region of $\omega$ Centauri, so it can improve the scenario that BSs in the low density cluster outskirts were formed by mass transfer in close binaries. NV358 is a second-generation binary, making it an evidence of multiple populations of the cluster $\omega$ Centauri. NV358 was discovered in the outer region but not in the center of $\omega$ Centauri, improving the result that $\omega$ Centauri was previously the nucleus of a nucleated dwarf galaxy and later it has been completely destroyed by a gravitational interaction between the Milky Way and the nucleated dwarf galaxy so that only its nucleus is now observed (\citealt{li12}). In the future, we hope to obtain the spectroscopic data, making it possible to evaluate more accurate physical parameters of the two binaries. This will provide evidence needed to improve our results.

\begin{figure}
\begin{center}
\includegraphics[angle=0,scale=1.1]{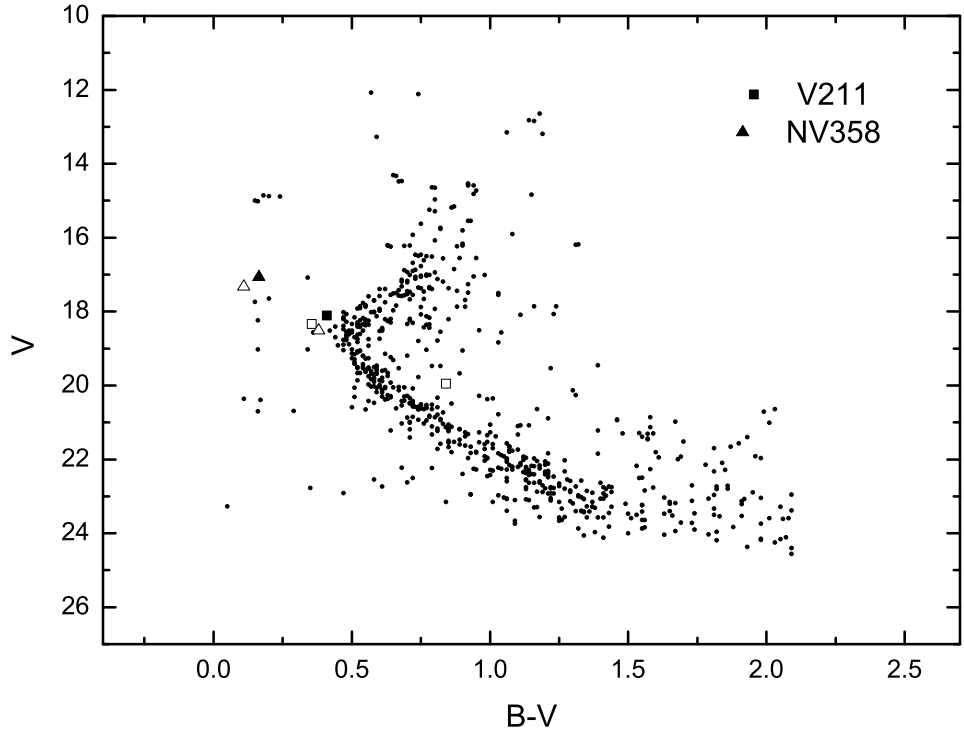}
\caption{Positions of the two binaries in the CMD for $\omega$ Centauri. The solid square gives the position of V211 and the solid triangle shows the position of NV358, open symbols represent the positions of the respective components.}
\end{center}
\end{figure}

\normalem
\begin{acknowledgements}
This work is partly supported by Chinese Natural Science Foundation
(Nos. 11203066, 11133007, 11103074, 10973037 and 10903026) and by
the West Light Foundation of the Chinese Academy of Sciences.
Thanks for the photometric data published in Kaluzny et al. (2004).

\end{acknowledgements}

\label{lastpage}

\end{document}